\documentclass[12pt]{article}
\usepackage{graphicx}
\usepackage{xcolor}
\usepackage[square,sort,comma,numbers]{natbib}

\def\institute{Department of Physics\\
Florida State University, Tallahassee FL 32306-4350, USA}
\def\support{\footnote{Work supported in part by the U.S. Department of Energy under Grant No. DE-SC0010102.}}

\def\Title#1{\begin{center} {\Large #1 } \end{center}}
\def\Author#1{\begin{center}{ \sc #1} \end{center}}
\def\Address#1{\begin{center}{ \it #1} \end{center}}

\newenvironment{Abstract}{\begin{quotation}  }{\end{quotation}}
\newenvironment{Presented}{\begin{quotation} \begin{center} 
             PRESENTED AT\end{center}\bigskip 
      \begin{center}\begin{large}}{\end{large}\end{center} \end{quotation}}
\def\Acknowledgements{\bigskip  \bigskip \begin{center} \begin{large}
             \bf ACKNOWLEDGEMENTS \end{large}\end{center}}

\def\beq{\begin{equation}}
\def\eeq#1{\label{#1}\end{equation}}
\def\eeqn{\end{equation}}

\def\beqa{\begin{eqnarray}}
\def\eeqa#1{\label{#1}\end{eqnarray}}
\def\eeqan{\end{eqnarray}}

\let\bar=\overbar

\def\Dslash{\not{\hbox{\kern-4pt $D$}}}
\def\dslash{\not{\hbox{\kern-2pt $\del$}}}

\def\msb{{\bar{\ssstyle M \kern -1pt S}}}

\begin{document}
\begin{titlepage}

\vfill
\Title{QCD NLO corrections to $t\bar{t}Z$ production at the LHC including leptonic decays}
\vfill
\Author{Laura Reina\support}
\Address{\institute}
\vfill
\begin{Abstract}
  The calculation of $t\bar{t}Z$ production including leptonic decays
  of both $Z$ boson and top quarks via a partially or fully off-shell
  calculation represents a definite step forward towards a more
  accurate description of events whose signatures originate from
  $t\bar{t}Z$ production. In this talk we review two studies that
  represent the state of the art of theoretical modelling of
  $t\bar{t}Z$ production by either performing the NLO QCD calculation
  of the fully decayed
  $pp\rightarrow e^+\nu_e\mu^-\bar{\nu}_\mu b\bar{b}\tau^+\tau^-$
  process~\cite{Bevilacqua:2022nrm} or matching the NLO QCD calculation of
  $pp\rightarrow t\bar{t}\ell^+\ell^-$ to parton shower, including top-quark
  decays with approximate LO spin correlation~\cite{Ghezzi:2021rpc}.
\end{Abstract}
\vfill
\begin{Presented}
$15^\mathrm{th}$ International Workshop on Top Quark Physics\\
Durham, UK, 4--9 September, 2022
\end{Presented}
\vfill
\end{titlepage}
\def\thefootnote{\fnsymbol{footnote}}
\setcounter{footnote}{0}

\section{Introduction}
\label{sec:intro}

Top-quarks, the most massive elementary fermions predicted by the
Standard Model (SM), are being copiously produced at the Large Hadron
Collider (LHC) and an unprecedented number of top-quark events will be
accumulated during the LHC lifetime including its high-luminosity
upgrade (HL-LHC). Rare processes where top quarks (either single or
pairs of) are produced in association with electroweak (EW) gauge
bosons ($W,Z,\gamma$) and the Higgs boson ($H$) will be accurately
measured, and will test the consistency of the top-quark EW and Yukawa
couplings with SM predictions. Processes consisting of top-quarks and
EW gauge bosons also constitute irreducible backgrounds to signatures
of physics beyond the SM (BSM) and their modelling needs to be well
understood in order to enhance the LHC discovery potential.

The reach of the HL-LHC top-quark physics program will remain the
state of the art in top-quark physics for quite some time, sice only
$e^+e^-$ colliders with energies above 500 GeV or very high-energy
hadron colliders (70-100 TeV or higher) will be able to substantially
improve on its results. It is therefore quite compelling to enable the
full top-quark physics reach of the LHC and HL-LHC by providing the
level of theoretical tools needed to adequately interpret existing and
future measurements. This requires considering the complexity of
signatures of various top-quark production processes and the effect of
restricting to specific fiducial volumes (used in experimental
analyses) when QCD and EW corrections are included and the complexity
of hadronic events is modelled by parton-shower evolution.

Here we consider $t\bar{t}Z$ production, a portal to access anomalies
in the top-quark coupling to the $Z$ boson and one of the main
backgrounds to $t\bar{t}H$ production as well as several BSM
searches. Studies of LHC data have already obtained constraints on
anomalous top-quark couplings to the $Z$ boson either in terms of form
factors that modify the SM vector and axial-vector $t\bar{t}Z$
couplings and add new tensor-like couplings~\cite{CMS:2019too} or in
terms of families of effective interactions that systematically extend
the SM Lagrangian and affect multiple top-quark
observables~\cite{Miralles:2021dyw}.

On the theoretical side, predictions for on-shell $t\bar{t}Z$
production including either next-to-leading (NLO) QCD
corrections~\cite{Lazopoulos:2008de,Kardos:2011na,Maltoni:2015ena} or
both NLO QCD and EW corrections~\cite{Frixione:2015zaa} have been
available for quite some time.  The NLO QCD calculation of on-shell
$t\bar{t}Z$ production has also been interfaced with parton-shower in
the \texttt{Powhel} framework~\cite{Garzelli:2011is,Garzelli:2012bn}
and independently made available in several public frameworks such as
\texttt{MG5\_aMC\@NLO}, \texttt{Sherpa}, and \texttt{POWHEG
  BOX}. Improved on-shell predictions also including resummation of
large soft logarithms up to next-to-next-to-leading logarithms (NNLL)
have been presented in multiple
studies~\cite{Broggio:2017kzi,Kulesza:2018tqz,Broggio:2019ewu,Kulesza:2020nfh}. Comparison
of the NLO+NNLL total cross section for $t\bar{t}Z$ on-shell
production~\cite{Broggio:2019ewu} with the most recent ATLAS and CMS
results~\cite{ATLAS:2021fzm,CMS:2019too} shows consistency within the
statistical and systematic errors which are currently at or above the
10\% level.

Moving forward, a more accurate comparison of not only total but in
particular differential distributions will be mandatory. Effective
interactions from BSM physics will most likely change the shape of
distributions and affect in particular their high-energy tails and
end-point regions.  Furthermore, experiments reconstruct final states
made of the decay products of $t\bar{t}Z$. A closer look to the
specific of LHC measurements of $t\bar{t}Z$ production shows that both
ATLAS~\cite{ATLAS:2021fzm} and CMS~\cite{CMS:2019too} have used
signatures containing three ($3\ell$) and four leptons ($4\ell$) plus
jets, including up to two $b$ jets, and missing energy, as arising
from $t\bar{t}Z$ production when the $Z$ boson decays leptonically
($Z\rightarrow \ell^+\ell^-$, $\ell=e,\mu,\tau$) and the $W^\pm$
bosons from (anti)top-quark decay leptonically or to light quarks
($W^\pm\rightarrow\ell^\pm\,\nu^{\!\!\!\!\!\!\!(-)},
q\bar{q}^\prime$). Differential cross sections based on the kinematic
of final-state leptons are most often used to analyze data and
identify the $t\bar{t}Z$ signal/background and most recently studies
based on final state signatures connected to multiple production
processes have been used to constrain classes of effective
interactions~\cite{CMS:2020lrr, CMS:2022hjj}.

In the attempt to reduce the theoretical systematic from modelling of
$t\bar{t}Z$ events, recent theoretical studies have focused on
signatures from fully or partially decayed $t\bar{t}Z$ production when
NLO QCD corrections are taken into account and the final state is
interfaced with a parton-shower event generator. In particular,
Ref.~\cite{Bevilacqua:2022nrm} has presented the fully off-shell
production of
$pp\rightarrow e^+\nu_e\mu^-\bar{\nu}_\mu b\bar{b}\tau^+\tau^-$
including fixed-order NLO QCD corrections, while
Ref.~\cite{Ghezzi:2021rpc} has studied
$pp\rightarrow t\bar{t}\ell^+\ell^-$ production accounting for
top-decays with approximate LO spin correlation and interfacing the
fixed-order NLO QCD calculation with parton shower using the POWHEG
method implemented in the publicly available \texttt{POWHEG BOX-V2}
framework. Highlights from both studies are presented in
Sections~\ref{sec:off-shell} and \ref{sec:ttee} respectively.

\section{NLO QCD
  \boldmath$pp\rightarrow e^+\nu_e\mu^-\bar{\nu}_\mu
  b\bar{b}\tau^+\tau^-$\unboldmath }
\label{sec:off-shell}

\begin{figure}[htb]
\centering
\includegraphics[scale=0.85]{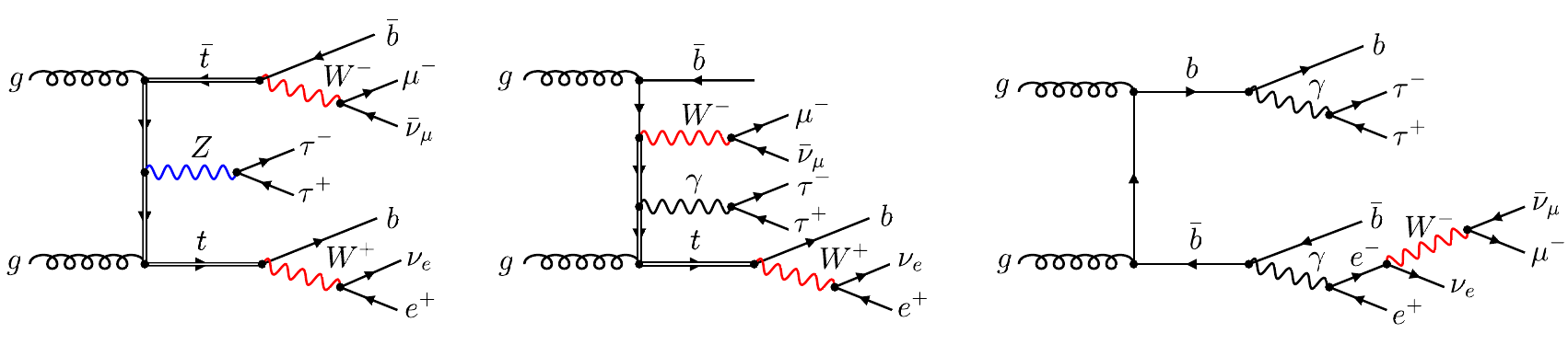}
\caption{Examples of double-resonant, single-resonant, and non-resonant
  parton-level processes contributing to
  $pp\rightarrow e^+\nu_e\mu^-\bar{\nu}_\mu b\bar{b}\tau^+\tau^-$.}
\label{fig:off-shell}
\end{figure}
The fully off-shell production of
$pp\rightarrow e^+\nu_e\mu^-\bar{\nu}_\mu b\bar{b}\tau^+\tau^-$
consists of double resonant, single resonant, and non-resonant
contributions as illustrated in Fig.~\ref{fig:off-shell}, where the
top quark, photon, $W^\pm$, and $Z$ gauge bosons are
off-shell. Ref.~\cite{Bevilacqua:2022nrm} has studied the impact of
all off-shell effects in a fully NLO QCD calculation of the above
process obtained within the \texttt{HELAC-NLO}\cite{Bevilacqua:2011xh}
framework. This automatically includes non factorizable QCD
corrections and QCD corrections to top-quark decay which are normally
omitted in the NLO QCD calculation of the corresponding signature from
on-shell $t\bar{t}Z$ production.  Results have been presented for
$\sqrt{s}=13$~TeV and using a fiducial volume defined by the following
conditions on the transverse momentum, rapidity, and
rapidity-azimuthal-plane distance among final state particles:
$p_T^{\ell}>20$~GeV, $|y^{\ell}|<2.5$, and $\Delta R_{\ell\ell}>0.4$
for leptons; $p_T^{b}>25$~GeV, $|y^{b}|<2.5$, and $\Delta R_{bb}>0.4$
for $b$ jets; and missing transverse momentum
$p_T^{\mathrm{miss}}>40$~GeV.

At NLO QCD the residual systematic uncertainty due to the choice of
renormalization and factorization scales has been estimated by using
both a fixed ($\mu=m_t+m_Z/2$) and a dynamical scale ($\mu=H_T/3$
where $H_T=\sum_i p_{T,i}$ summed over the final-state particles and
missing momentum) with a traditional 7-point variation, and found to
be as low as 6\% for the total cross section and within 10\% for most
distributions. While both scale choices produce compatible results at
the level of the total cross section, the dynamical scale proves to be
a better choice at the level of distributions.  As part of the
residual theoretical uncertainty, the dependence on the choice of
parton distribution functions (PDF) obtained by comparing results that
use different PDF sets has been studied and determined to be within
1-3\%.
\begin{figure}[htb]
\centering
\includegraphics[scale=0.4]{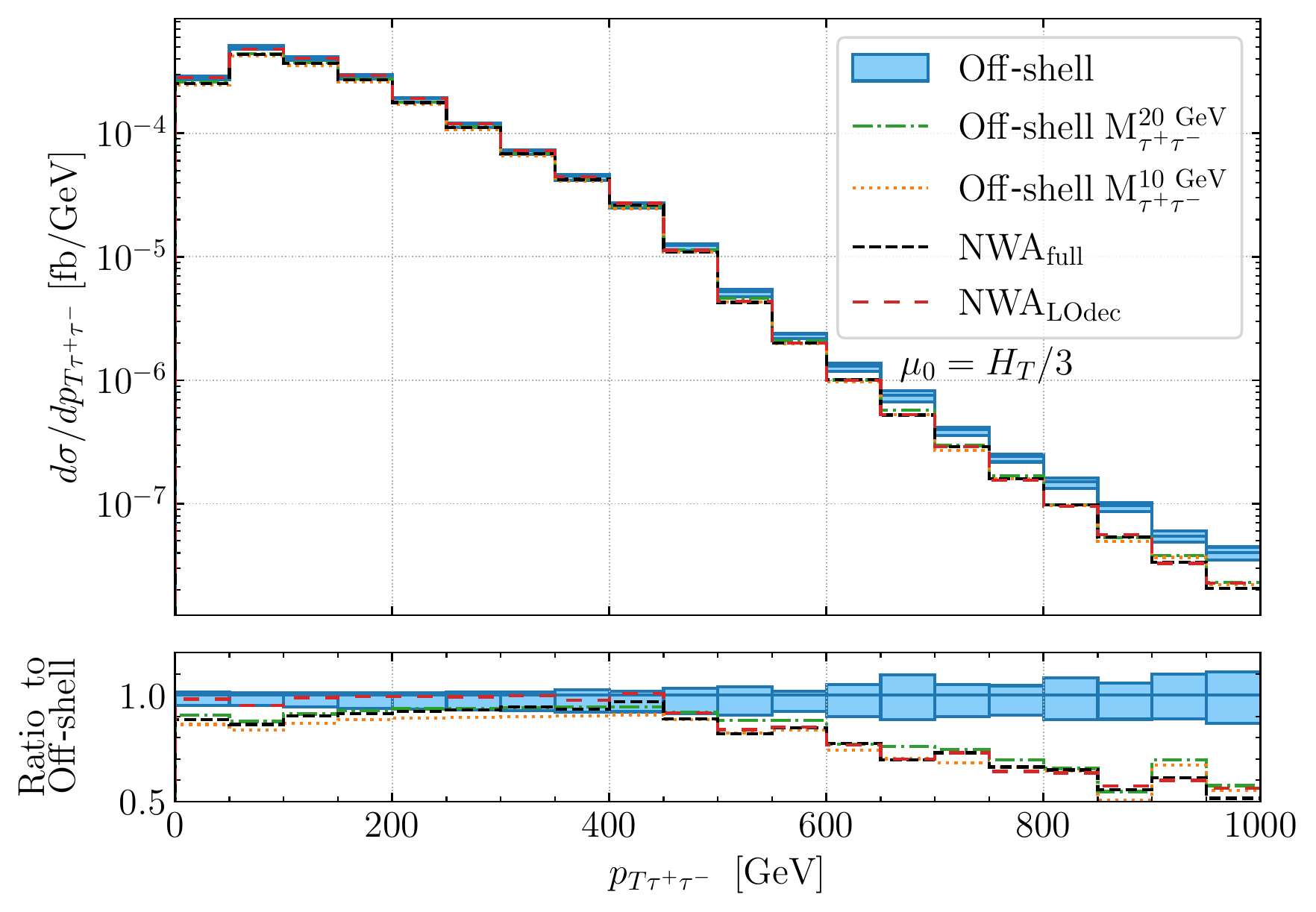}
\includegraphics[scale=0.4]{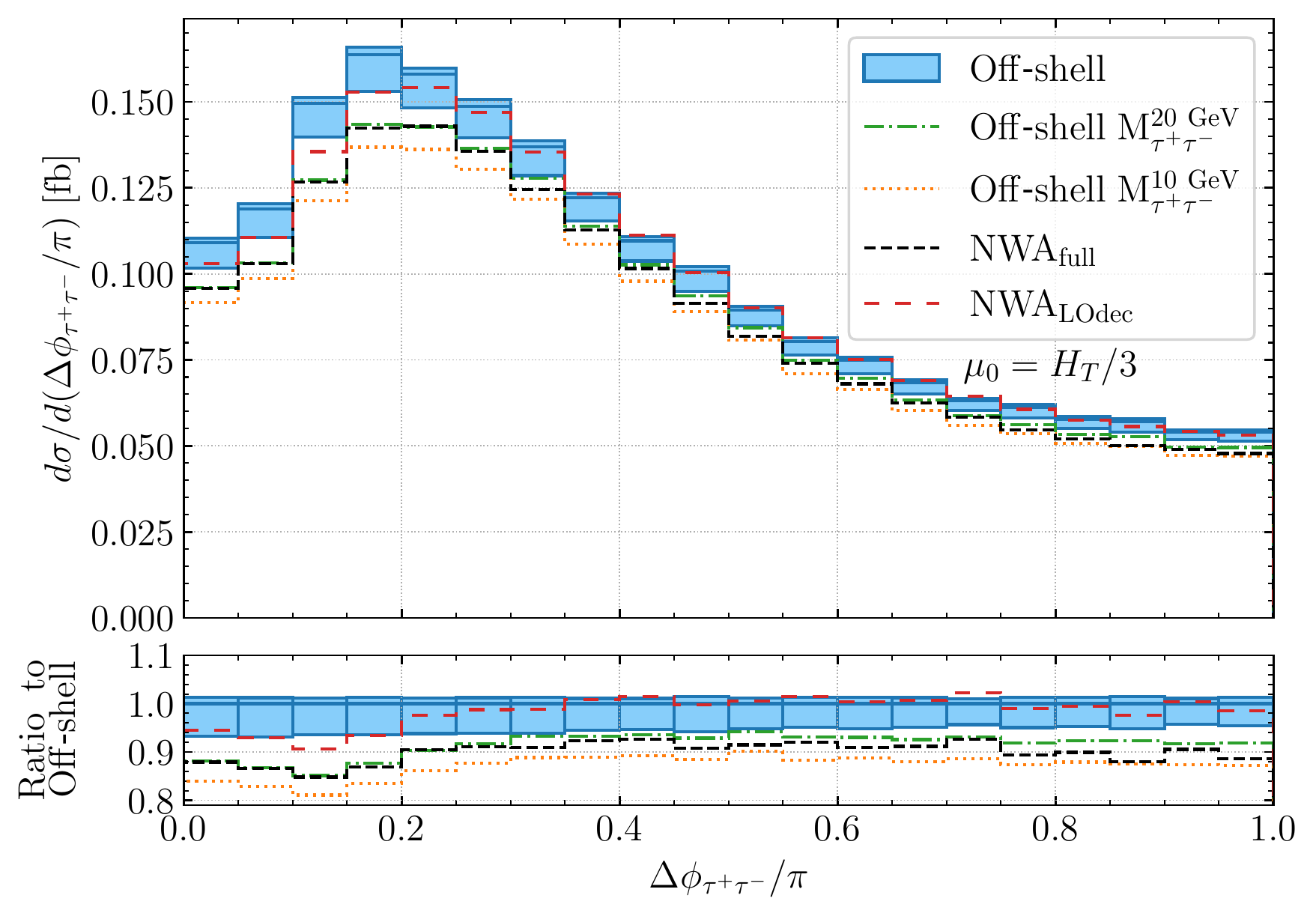}
\caption{Differential cross section as a function of
  $p_{T,\tau^+\tau^-}$ and $\Delta\phi_{\tau^+\tau^-}$ calculated in the
  full off-shell or NWA approximation, with LO or NLO top-quark
  decays. From Ref.~\cite{Bevilacqua:2022nrm}.} 
\label{fig:dist-off-shell}
\end{figure}

NLO QCD corrections affect different regions of kinematic
distributions differently and cannot be mimicked by an overall
rescaling factor. Furthermore, NLO QCD corrections to top-quark decay
are relevant and represent almost 9\% of the full corrections. A
comparison with results obtained in the Narrow Width Approximation
(NWA) shows how off-shell effects are most prominent in tails of
distributions and in the vicinity of kinematic end points where a NWA,
with NLO QCD description of top-quark decays, ceases to be a valid
approximation of the full off-shell calculation. Prototype examples to
illustrate these results are the transverse momentum ($p_T$) and
azimuthal separation ($\Delta\phi)$) distributions of the
$\tau^+\tau^-$ system shown in Fig.~\ref{fig:dist-off-shell}. A
complete set of results can be found in
Ref.~\cite{Bevilacqua:2022nrm}.

\section{NLO QCD \boldmath$pp\rightarrow t\bar{t}\ell^+\ell^-$\unboldmath
  matched to parton shower}
\label{sec:ttee}

\begin{figure}[htb]
\centering
\includegraphics[scale=0.6]{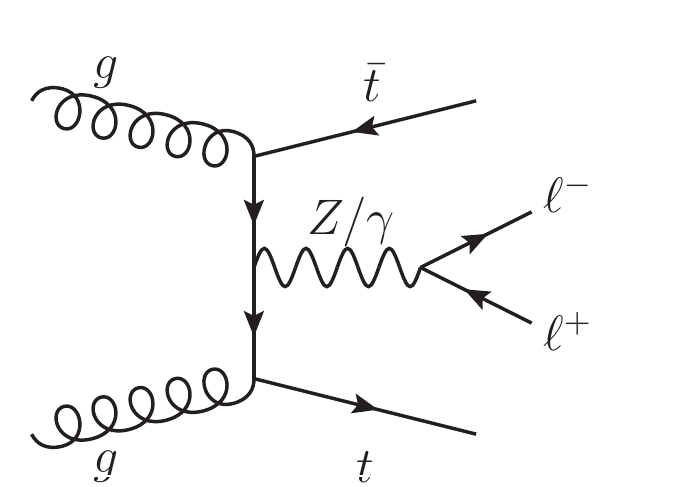}
\includegraphics[scale=0.6]{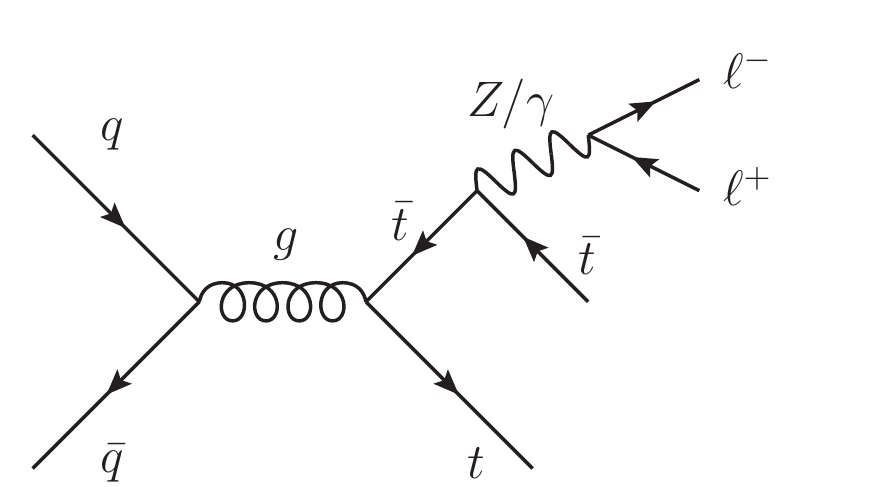}
\caption{Parton-level channels for $pp\rightarrow t\bar{t}\ell^+\ell^-$.}
\label{fig:ttll}
\end{figure}
Accounting for QCD effects in parton evolution is also crucial to
reach higher accuracy in modelling hadron-collider events. As a first
step towards the interface of the NLO QCD calculation of fully
off-shell signatures of $t\bar{t}Z$ production with parton shower,
Ref.~\cite{Ghezzi:2021rpc} has considered the
$pp\rightarrow t\bar{t}\ell^+\ell^-$ process and interfaced it with
parton-shower by matching with \texttt{PYTHIA} using the POWHEG
method~\cite{Nason:2004rx, Frixione:2007vw} within the \texttt{POWHEG
  BOX-V2} framework~\cite{Alioli:2010xd}. The one-loop matrix elements
have been obtained via \texttt{NLOX}~\cite{Honeywell:2018fcl,
  Figueroa:2021txg}, while decays of the top quarks have been modelled
by keeping LO spin correlation~\cite{Frixione:2007zp} without
accounting for QCD corrections. As illustrated by the tree-level
processes depicted in Fig.~\ref{fig:ttll}, the $t\bar{t}\ell^+\ell^-$
signature can be produced both via $Z$ or photon ($\gamma$)
exchange. As noticed also in Ref.~\cite{Bevilacqua:2022nrm}, the
photon's contribution (both direct and through the interference with
the $Z$'s one) is a substantial part of the off-shell effects and can
be reduced by restricting the invariant mass of the $\ell^+\ell^-$
pair to lie within a narrow window centered around $m_Z$,
i.e. imposing a window cut of the form
$m_Z-\Delta<m_{\ell^+\ell^-}<m_Z+\Delta$. The results of
Ref.~\cite{Ghezzi:2021rpc} uses $\Delta=10$ GeV to mimic choices
adopted in recent experimental analyses~\cite{CMS:2019too,
  ATLAS:2021fzm}. Results are presented for $\sqrt{s}=13$~TeV and
imposing the following additional cuts on the final-state leptons:
$p_T^{\ell}>10>$~GeV and $|y^{\ell}|<2.5$.

Ref.~\cite{Ghezzi:2021rpc} estimates the residual theoretical
systematic uncertainty from renormalization and factorization scale
dependence by comparing both a fixed ($\mu=m_t+m_Z/2$) and a dynamical
($(M_T(t)+M_T(\bar{t})+M_T(\ell^+\ell^-))/3$, with
$M_T(i)=\sqrt{m_i^2+p_{T,i}^2}$) scale varied by a factor of two
according to a 7-point variation, and find it to be around 10\% for
both total and differential $t\bar{t}\ell^+\ell^-$ cross sections
independently of the kind of scale used. As for the fully off-shell
case discussed in Sec.~\ref{sec:off-shell}, NLO QCD corrections
modifies the shape of distributions and are a mandatory component of
modelling these signatures, even more so when matching to
parton shower.

\begin{figure}[htb]
\centering
\includegraphics[scale=0.8]{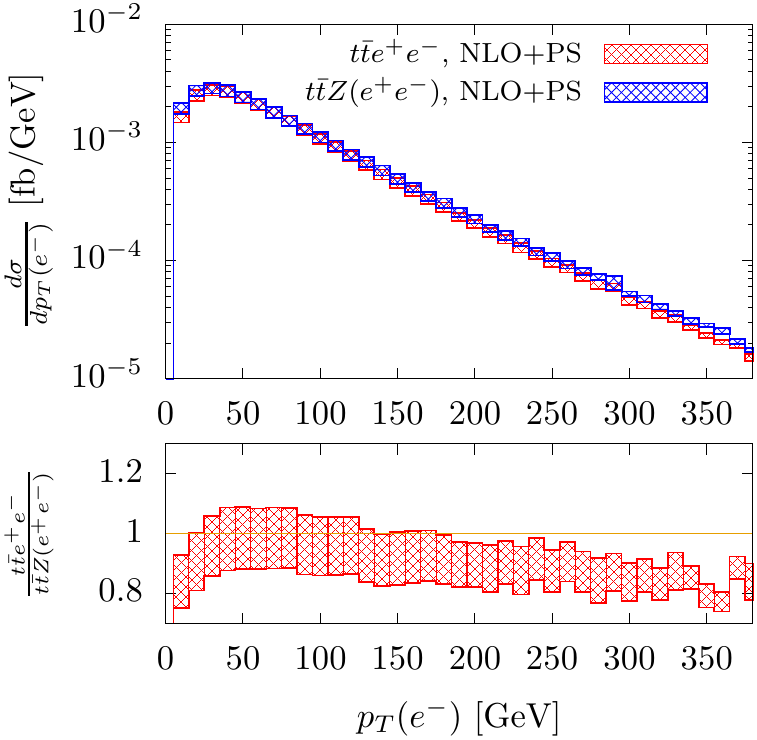}
\hspace{1.truecm}
\includegraphics[scale=0.85]{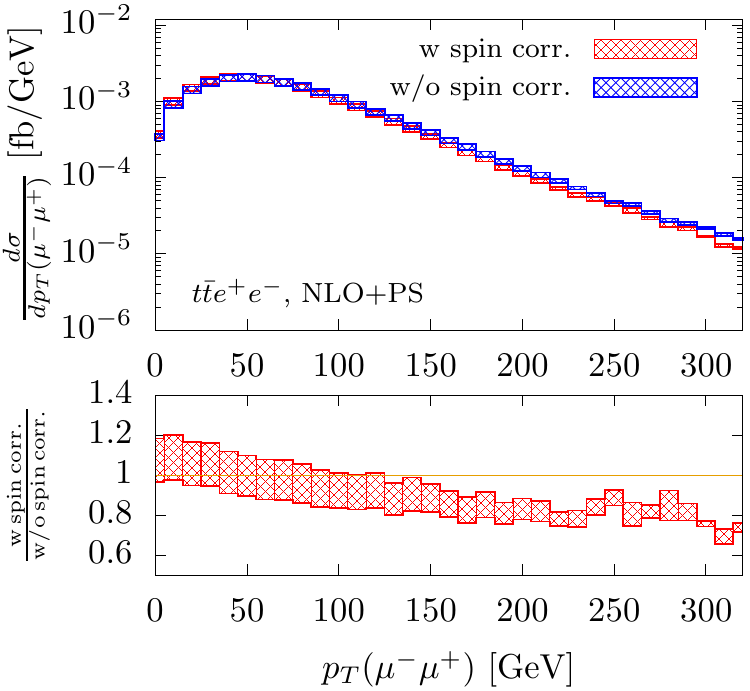}
\caption{Differential cross section as a function of
  $p_T(e^-)$ (l.h.s.) and $p_T(\mu^+\mu^-)$ (r.h.s) at NLO QCD
  interfaced with parton shower. On the l.h.s. the
  top-quarks are not decayed, and the comparison with on-shell
  $t\bar{t}Z$ production is shown. The r.h.s. shows the effect of decaying
  the top quarks without accounting for spin correlation or accounting
  for it through LO matrix elements. From Ref.~\cite{Ghezzi:2021rpc}.}
\label{fig:dist-ttll}
\end{figure}

The l.h.s. of Fig.~\ref{fig:dist-ttll} illustrates the comparison of
on-shell and off-shell effects considering $t\bar{t}e^+e^-$ production
as calculated via the $t\bar{t}\ell^+\ell^-$ process or via
$t\bar{t}Z$ on-shell production with the $Z$ decaying into $e^+e^-$
via the parton shower. For this comparison the top quarks are not
decayed. Effects of 10-20\% from off-shell $Z$ and spin correlation of
the decay products are clearly visible in the tail of the $p_T(e^-)$
distribution even for relatively moderate $p_T$. These effects are
visible on top of the current theoretical systematics from scale
uncertainty and even after interfacing with parton-shower. Switching
on the decay of the top quarks (here taken to be
$t\rightarrow b \mu^+\nu_\mu$ and
$\bar{t}\rightarrow \bar{b} \mu^-\bar\nu_\mu$) allows to investigate
the effect of including spin correlations, even if only with LO matrix
elements and in a NWA. This is illustrated in the r.h.s. plot of
Fig.~\ref{fig:dist-ttll} which shows the $p_T$ distribution of the
$\mu^+\mu^-$ pair. Spin correlation effects are clearly visible on top
of the theoretical systematic uncertainty. Similar effects are present
in adimensional distributions such as rapidity and relative
azimuthal-angle distribution. A complete set of results can be found
in Ref.~\cite{Ghezzi:2021rpc}.

\section{Conclusions}
\label{sec:concl}

Enabling the top-quark precision program of the (HL-)LHC is a priority
given the relevance this can have in starting to answer fundamental
open questions such as the origin of the EW scale and the hierarchy of
fermion masses. In this context, processes where top quarks are
produced with $X=W^\pm,Z,\gamma$, and $H$ play a crucial role in
constraining BSM physics that can affect multiple top-quark
couplings. Deviations are expected to be more visible in kinematical
regions, such as tails and end points of distributions, that can be
limited by the reach of the machine and require a focused effort on
both the experimental and theoretical side. Theoretical accuracy and
flexibility in modelling the complexity of LHC events will be crucial.
Important steps in this direction have been accomplished by presenting
a fully off-shell calculation of
$pp\rightarrow e^+\nu_e\mu^-\bar{\nu}_\mu b\bar{b}\tau^+\tau^-$ at NLO
QCD in Ref.~\cite{Bevilacqua:2022nrm} and by interfacing the NLO QCD
calculation of $t\bar{t}\ell^+\ell^-$ with a parton-shower in
Ref.~\cite{Ghezzi:2021rpc}. Further developments should include the
interface of the fully off-shell calculation with parton shower and
possibly accounting for higher order QCD corrections when results
for fixed-order NNLO QCD effects in $t\bar{t}Z$ production
become available.

\Acknowledgements The work of L.R. is supported in part by the
U.S. Department of Energy under grant DE-SC0010102. She would also
like to thank the organizers of TOP 2022 and the Florida State
University College of Arts and Sciences for supporting her travel
expenses, as well as M. Ghezzi, B. J\"ager, S. Lopez Portillo Chavez,
and D. Wackeroth for a pleasant and fruitful collaboration.

\end{document}